\documentclass[aps,preprint,showpacs,superscriptaddress,groupedaddress]{revtex4}  
\usepackage{graphicx}  
\usepackage{dcolumn}   
\usepackage{bm}        
\usepackage{amssymb}   
\usepackage{amsmath}
\usepackage{amsfonts}

\begin{document}



\title{Multipole Expansion in Generalized Electrodynamics}
\affiliation{Federal University of Technology - Paran\'{a} - C\^{a}mpus Ponta Grossa - Av. Monteiro Lobato, s/n - Km 04 - CEP 84016-210 - Ponta Grossa - Brazil\\%
carlosbonin@utfpr.edu.br (corresponding author)}
\affiliation{Institute for Theoretical Physics - S\~{a}o Paulo State University (IFT/UNESP), Rua Dr. Bento Teobaldo Ferraz 271 - Bl. II - Barra Funda CEP 01140-070 - S\~{a}o Paulo, S\~{a}o Paulo - Brazil\\%
pimentel@ift.unesp.br }
\affiliation{Universidade Federal do Rio de Janeiro, Reitoria, Centro de Tecnologia. CT - Centro de Tecnologia Cidade Universit\'{a}ria 21941909 - Rio de Janeiro, RJ - Brazil - Caixa-postal: 68528\\%
ortega@if.ufrj.br}
\author{C. A. Bonin} \affiliation{Federal University of Technology - Paran\'{a} - C\^{a}mpus Ponta Grossa - Av. Monteiro Lobato, s/n - Km 04 - CEP 84016-210 - Ponta Grossa - Brazil\\%
carlosbonin@utfpr.edu.br (corresponding author)}
\author{B. M. Pimentel} \affiliation{Institute for Theoretical Physics - S\~{a}o Paulo State University (IFT/UNESP), Rua Dr. Bento Teobaldo Ferraz 271 - Bl. II - Barra Funda CEP 01140-070 - S\~{a}o Paulo, S\~{a}o Paulo - Brazil\\%
pimentel@ift.unesp.br }
\author{P. H. Ortega} \affiliation{Physics Institute - Federal University of Rio de Janeiro (IF-UFRJ),
Av. Athos da Silveira Ramos 149 - Centro de Tecnologia, Bloco A, 3o Andar, Cidade Universit\'{a}ria - CEP: 21941-909 - Caixa Postal: 68528 - Rio de Janeiro, Rio de Janeiro - Brazil\\%
ortega@if.ufrj.br}

\date{\today}

\begin{abstract}
In this article we study some classical aspects of Podolsky Electrodynamics in the static regime. We develop the multipole expansion for the theory in both the electrostatic and the magnetostatic cases. We also address the problem of consistently truncating the infinite series associated with the several kinds of multipoles, yielding approximations for the static Podolskian electromagnetic field to any degree of precision required. Moreover, we apply the general theory of multipole expansion to some specific physical problems. In those problems we identify the first terms of the series with the monopole, dipole and quadrupole terms in the generalized theory. We also propose a situation in which Podolsky theory can be experimentally tested.
\end{abstract}

\maketitle


\section{Introduction}
Rare are the physical problems to which we can provide an exact solution. Despite some of those give us insights into more complex systems, as it is the case of the so-called \textit{toy models} used in many areas, and very few can be directly applied to more general cases, \textit{e. g.} the harmonic oscillator, most of those problems are just too simple to describe the majority of the observed phenomena. For other cases, the most accurate approach consists in trying to find an approximated solution to the problem. Examples of these cases are the problems we deal with in the present paper. Our goal is to investigate the electrostatics and magnetostatics of the so-called Generalized Electrodynamics. Generalized Electrodynamics is an alternative theory developed by Podolsky in the nineteen-forties with the hope of getting rid of the divergencies present in the Maxwellian Quantum Electrodynamics \cite{Podolsky}. Although Podolsky appeared to have failed in that goal, many years later it was shown by one of us and a collaborator that Podolsky did not use a proper gauge fixing condition in his theory, leading to spurious results \cite{Pimentel}. Among the various features of Podolsky theory, we cite the fact that it is a theory with higher-order derivatives and it depends on a free parameter, namely, the Podolsky parameter $m_P$. The best estimate for this parameter, calculated by one of us and collaborators, sets it to be $m_P>3.7595\times 10^{10}eV$ \cite{Bufalo}. It was shown that Podolsky theory is the only linear, local extension of Maxwell theory with second-order derivatives which is both Poincar\'e and gauge $U(1)$-invariant \cite{Cuzinatto}. Furthermore, Generalized Quantum Electrodynamics has been studied in thermodynamic equilibrium by two of us \cite{Bonin} and have also shown that, in the case of free quantum electromagnetic field in thermodynamic equilibrium, Podolsky theory results in a deviation of Stefan-Boltzmann law \cite{Stefan Boltzmann}.

Despite many recent advances towards the understanding of Generalized Quantum Electrodynamics, its classical counterpart remains only superficially studied. In this regime, one of the most important results was the solution of the famous ``4/3 problem in Classical Electrodynamics" within the context of Podolsky theory \cite{Frenkel}. The possibility of probing Podolsky Electrodynamics through several experiments has been studied by Cuzinatto \textit{et. al.} \cite{Probing}. One of the most basic procedures in Maxwell Classical Electrostatics and Magnetostatics is the multipole expansion \cite{Jackson}. Such an expansion, as far as we know, has not been studied in the literature within the context of Podolsky theory and its development is the main goal of the present paper. The importance of such an expansion is that, in general, it is not possible to compute exactly the electric field generated by a static electric charge distribution (or the magnetic field due to static electric currents). In the Maxwellian theory, we appeal to an approximation to the Green Function of the field in terms of powers of the ratio between the typical size of the electric charge (or current) distribution and the distance between the point in which we want to evaluate the field and the distribution. In the Podolsky case, however, the situation is more involved. Due to the Podolsky parameter, besides the other typical lengths already present in the usual Maxwellian case, a new physical length appears, namely, $l_P = 1/\left|m_P\right|$. For this reason, the multipole expansion in the generalized theory is not at all straight-forward. In this article, following the Master Degree Dissertation of one of us \cite{Paulo}, we carefully develop the multipole expansion in Podolsky theory in both the electrostatic and magnetostatic regimes. We also apply our results to some classical problems of electrostatics, and magnetostatics as well.


\section{The Podolsky Equations in the Static Case}

The Action Functional $S$ for the Generalized Electrodynamics in a four-dimensional Minkowski spacetime with a metric $\eta$ such that $diag\left(\eta\right)=\left(1,-1,-1,-1\right)$ is

\begin{align}
S &=\int_\Omega d^4x \,\mathcal{L}_P\left(A,\partial A, \partial^2 A; J\right),
\end{align}
where $\Omega$ is the (simply-connected) spacetime region for the problem, $\mathcal{L}_P$ is the Podolsky Lagrangian Density, $A$ is the electromagnetic field, and $J$ is the field source. Unlike Maxwell Lagrangian Density, Podolsky Lagrangian Density depends on both first and second-order derivatives of the electromagnetic field. $\mathcal{L}_P$ has the form\footnote{Throughout this text we use the Einstein summation convention: repeated Greek indexes are summed from 0 to 3, latin indexes are summed over from 1 to 3, and we use the natural unit system as well unless explicitly stated otherwise.} \cite{Podolsky}

\begin{align}
\mathcal{L}_P &= -\frac{1}{4}F^{\mu\nu}F_{\mu\nu}+\frac{1}{2m_P^2}\partial_\mu F^{\mu\xi}\partial^{\nu}F_{\nu\xi}-J^\mu A_\mu.
\end{align}

In this expression the field-strength tensor $F$ has components $F_{\mu\nu}=\partial_\mu A_\nu - \partial_\nu A_\mu$ and $m_P$, as mentioned in the Introduction, is a nonvanishing (henceforth, without loss of generality, assumed positive) parameter with dimension of energy called the \textit{Podolsky parameter}. The Euler-Lagrange equations obtained from the action $S$ are

\begin{align}
\left(\frac{\square}{m_P^2}+1\right)\partial_\mu F^{\mu\nu} &= J^\nu.\label{EL}
\end{align}
Here, $\square = \partial^\mu\partial_\mu$ stands for the d'Alembertian operator. As it is clear from the definition of the field-strength, the Bianchi identity is also satisfied:

\begin{align}\label{Bianchi}
\partial_\mu F_{\nu\xi} + \partial_\nu F_{\xi\mu} + \partial_\xi F_{\mu\nu} &=0.
\end{align}

We can also rewrite the eight nontrivial equations (\ref{EL}) and (\ref{Bianchi}) in terms of the electric and the magnetic fields $\mathbf{E}$ and $\mathbf{B}$, whose components are given by

\begin{align}
E^i &= - F^{0i};\\
B^i &= -\frac{1}{2}\epsilon^{ijk}F_{jk},
\end{align}
where $\epsilon$ is the Levi-Civita symbol of rank 3, as

\begin{align}
\left(\frac{\square}{m_P^2}+1\right)\nabla\cdot\mathbf{E} &= \rho;\\
\nabla\cdot\mathbf{B} &=0;\\
\left(\frac{\square}{m_P^2}+1\right)\left(\nabla\times\mathbf{B}-\frac{\partial\mathbf{E}}{\partial t}\right) &= \mathbf{j};\\
\nabla\times\mathbf{E}+\frac{\partial\mathbf{B}}{\partial t} &=\mathbf{0}.
\end{align}

In these equations we have used the wide-spread notation $J=\left(\rho,\mathbf{j}\right)$ in which $\rho$ represents the electric charge density and $\mathbf{j}$ the electric current surface density.

In the static case, \textit{i. e.}, the case in which neither the electric nor the magnetic fields depend on time, Podolsky equations take the simplified form

\begin{align}
\left(1-\frac{\nabla^2}{m_P^2}\right)\nabla\cdot\mathbf{E} &= \rho;\\
\nabla\cdot\mathbf{B} &=0;\\
\left(1-\frac{\nabla^2}{m_P^2}\right)\nabla\times\mathbf{B} &= \mathbf{j};\\
\nabla\times\mathbf{E} &=\mathbf{0},
\end{align}
where the electric charge and current surface densities are time-independent as well.

Just like in the usual case, we can write the static electric and magnetic fields in terms of the components of the time-independent electromagnetic field $A=\left(\varphi,\mathbf{A}\right)$, with $\varphi$ being the electrostatic (also known as scalar) potential and $\mathbf{A}$ the so-called vector potential:

\begin{align}
\mathbf{E} &= -\nabla\varphi;\\
\mathbf{B} &= \nabla\times\mathbf{A}.
\end{align}

Therefore, the potentials satisfy

\begin{align}
\left(\frac{\nabla^2}{m_P^2}-1\right)\nabla^2\varphi &= \rho;\label{phi}\\
\left(\frac{\nabla^2}{m_P^2}-1\right)\left[\nabla^2\mathbf{A}-\nabla\left(\nabla\cdot\mathbf{A}\right)\right] &= \mathbf{j}.\label{for A}
\end{align}

This last equation is a rather intricate one. However, just like all other gauge fields, Podolsky field presents more apparent degrees of freedom than physical ones. To deal with this discrepancy, we impose the \textit{generalized Coulomb condition} \cite{Pimentel}

\begin{align}
\left(\frac{\nabla^2}{m_P^2}-1\right)\nabla\cdot\mathbf{A} &=0.
\end{align}

With this condition, equation (\ref{for A}) becomes

\begin{align}
\left(\frac{\nabla^2}{m_P^2}-1\right)\nabla^2\mathbf{A} &= \mathbf{j}.\label{A}
\end{align}

The solution for equation (\ref{phi}) - or, for that matter, for any component of (\ref{A}) - has the form\footnote{In the case of the vector potential, $\rho$ is replaced by $\mathbf{j}$.}

\begin{align}
\varphi(\mathbf{r}) &= \int d^3 r'\, G\left(\mathbf{r},\mathbf{r}'\right)\rho\left(\mathbf{r}'\right),\label{potential}
\end{align}
where $G$ is the Green function of the operator $\left(\frac{\nabla^2}{m_P^2}-1\right)\nabla^2$:

\begin{align}
\left(\frac{\nabla^2}{m_P^2}-1\right)\nabla^2G\left(\mathbf{r},\mathbf{r}'\right) &=\delta\left(\mathbf{r}-\mathbf{r}'\right).\label{Green}
\end{align}

Here, $\delta\left(\mathbf{r}\right)$ is the Dirac delta function in three dimensions.

Since the operator $\left(\frac{\nabla^2}{m_P^2}-1\right)\nabla^2$ is translational invariant, its Green function depends only on the difference between the variables $\mathbf{r}$ and $\mathbf{r}'$: $G\left(\mathbf{r},\mathbf{r}'\right)=G\left(\mathbf{r}-\mathbf{r}'\right)$. Solving (\ref{Green}), we get

\begin{align}
G\left(\mathbf{r}\right) &=\frac{1-e^{-m_Pr}}{4\pi r}.\label{Green Function}
\end{align}
where we have written $\left|\mathbf{r}\right|=r$.

The physical content of this Green function is deep, if not apparent. First of all, for any fixed distance $r$, we can recover Maxwell's result in the limit

\begin{align}
\lim_{m_P\rightarrow\infty}G\left(\mathbf{r}\right)&=\frac{1}{4\pi r}.
\end{align}

Furthermore, the limit of infinitesimal distances of $G$ is quite surprising:

\begin{align}
\lim_{r\rightarrow 0^+}G\left(\mathbf{r}\right)=\frac{m_P}{4\pi}.
\end{align}

This result is related to the fact that the Podolsky's electrostatic field of a point electric charge, unlike Maxwell's, can be defined to be finite everywhere, including over the charge itself.

If we plug the Green function (\ref{Green Function}) into equation (\ref{potential}), \textit{in principle} we are able to compute the electrostatic potential due to the charge distribution $\rho$ at any point. We emphasize the words \textit{in principle} because there are very few charge distributions to which we can actually integrate $\int d^3 r'\, G\left(\mathbf{r}-\mathbf{r}'\right)\rho\left(\mathbf{r}'\right)$ in order to find the potential $\varphi$ at the point $\mathbf{r}$. One of those solvable cases takes place when the charge distribution is due to a set of $n$ point particles with electric charges $q_1$, $q_2$, ..., $q_n$ located at $\mathbf{r}_1$, $\mathbf{r}_2$, ..., $\mathbf{r}_n$, respectively:

\begin{align}
\rho\left(\mathbf{r}'\right)=\sum_{k=1}^n q_k\delta\left(\mathbf{r}'-\mathbf{r}_k\right).
\end{align}

For such a configuration, the electrostatic potential at the point $\mathbf{r}$ is found to be

\begin{align}
\varphi\left(\mathbf{r}\right)=\sum_{k=1}^n \frac{q_k}{4\pi\left|\mathbf{r}-\mathbf{r}_k\right|}\left(1-e^{-m_P\left|\mathbf{r}-\mathbf{r}_k\right|}\right).\label{discrete}
\end{align}

Unfortunately, for smoother configurations of charge, we are, in general, unable to compute the potential exactly and we need to perform approximations. One of the most successful approaches to compute the static potentials in the Maxwell theory for localised electric charges is the multipole expansion. Since the correspondent Maxwellian Green  function is simply $\left(4\pi\left|\mathbf{r}-\mathbf{r}'\right|\right)^{-1}$, the multipole expansion in Maxwell theory essentially consists in truncating the series

\begin{align}
\frac{1}{\left|\mathbf{r}-\mathbf{r}'\right|}=\frac{1}{r}\sum_{l=0}^\infty \left(\frac{r'}{r}\right)^l P_l\left(\cos\gamma\right),\label{approx}
\end{align}
where  $r'=\left|\mathbf{r}'\right|$ and the expansion is valid for $r'/r<1$, $P_l$ is the $l$-th Legendre Polynomial and $\gamma$ is the smallest angle between $\mathbf{r}$ and $\mathbf{r}'$.\footnote{The $l$-th Legendre Polynomial is $P_l\left(z\right)=\frac{1}{2^l l!}\frac{d^l}{dz^l}\left(z^2-1\right)^l$.} The physical meaning of the condition $r'/r<1$ is that we are computing the potential outside the region where the electric charge distribution is nonvanishing. For the equivalent expansion in the Podolsky theory, besides (\ref{approx}), we need \cite{ByronFuller}

\begin{align}
\frac{e^{-m_P\left|\mathbf{r}-\mathbf{r}'\right|}}{\left|\mathbf{r}-\mathbf{r}'\right|}=&\,\,m_P\sum_{l=0}^{\infty}\left(2l+1\right) i_l\left(m_Pr'\right) k_l\left(m_P r\right) P_l\left(\cos\gamma\right).
\end{align}

Here, $i_l$ and $k_l$ are the Modified Spherical Bessel Functions of the First and the Second Kinds, respectively, and order $l$.

With these expansions, we can write the Podolsky Static Green Function (\ref{Green Function}) as

\begin{align}
G\left(\mathbf{r},\mathbf{r}'\right)=&\frac{1}{4\pi r}\sum_{l=0}^{\infty}\left[\left(\frac{r'}{r}\right)^l-\left(2l+1\right)m_P r\,\, i_l\left(m_P r'\right) \, k_l\left(m_P r\right)\right]P_l\left(\cos\gamma\right).
\end{align}

By making use of the representation

\begin{align}
i_l\left(x\right)=\left(2x\right)^l\sum_{s=0}^\infty\frac{\left(s+l\right)!\,x^{2s}}{s!\left[2\left(s+l\right)+1\right]!},
\end{align}
$G$ can be rewritten as

\begin{align}
G\left(\mathbf{r},\mathbf{r}'\right)=\frac{1}{4\pi r}\sum_{l=0}^\infty G_l\left(\mathbf{r},\mathbf{r}'\right)\label{G Gl}
\end{align}
where

\begin{align}
G_l\left(\mathbf{r},\mathbf{r}'\right)=&\left(\frac{r'}{r}\right)^l P_l\left(\cos\gamma\right)-m_Pr\left(m_Pr'\right)^l\sum_{n=0}^{\tilde{l}}\frac{2^{\bar{n}}\left(2\bar{n}+1\right)\left(\tilde{l}-n+\bar{n}\right)!\,
k_{\bar{n}}\left(m_Pr\right)}{\left(\tilde{l}-n\right)!\left[2\left(\bar{n}+\tilde{l}-n\right)+1\right]!} P_{\bar{n}}\left(\cos\gamma\right)
\end{align}
and

\begin{align}
\bar{n}=&2n+\frac{1-\left(-1\right)^l}{2};\\
\tilde{l}=&\frac{2l-1+\left(-1\right)^l}{4}.
\end{align}

Since

\begin{align}
k_l\left(x\right)=\frac{e^{-x}}{x^{l+1}}\sum_{s=0}^{l}a_{s,l}x^s
\end{align}
for certain coefficients $a_{s,l}$, we see that in the regimes of distance\footnote{Notice that condition (\ref{ap 1}) is, in fact, a consequence of conditions (\ref{ap 2}) and ({\ref{ap 3}}).}

\begin{align}
\frac{r'}{r}&\ll 1;\label{ap 1}\\
m_P r'&\ll 1;\label{ap 2}\\
m_P r&\gtrsim 1, \label{ap 3}
\end{align}
the following condition holds:

\begin{align}
\frac{G_{l+1}\left(\mathbf{r},\mathbf{r}'\right)}{G_l\left(\mathbf{r},\mathbf{r}'\right)}\ll 1.
\end{align}

This condition is precisely the one needed in the multipole approximation.

Before we proceed, we must discuss the physical meaning of the relations (\ref{ap 1}-\ref{ap 3}). Condition (\ref{ap 1}) is the usual one encountered in the multipole approximation of Maxwell theory. It essentially states that the distance between the point where the potential is evaluated and the charge distribution must be much greater than the typical size of the charge distribution. The other two relations are novel to the traditional development and appear only in the Podolsky theory. The condition (\ref{ap 2}) states that the typical size of the nonvanishing charge distribution must be much smaller than the Podolsky length $l_P=1/m_P$. On the other hand, the relation (\ref{ap 3}) indicates that the distance between the point where the potential is computed and the charge distribution must be of the same order of magnitude of $l_P$ - or higher.

If we insert the series (\ref{G Gl}) into equation (\ref{potential}), we can write the multipole expansion for the static potential as

\begin{align}
\varphi\left(\mathbf{r}\right)=\sum_{l=0}^\infty\varphi^{\left(l\right)}\left(\mathbf{r}\right),\label{varphi}
\end{align}
where

\begin{align}
\varphi^{\left(l\right)}\left(\mathbf{r}\right)=\int d^3r'\frac{G_l\left(\mathbf{r},\mathbf{r}'\right)\rho\left(\mathbf{r}'\right)}{4\pi r}
\end{align}
is the so-called $2^l$-pole term of the potential.

In the regime of validity of relations (\ref{ap 1}-\ref{ap 3}), the dominant term in the series above is the \textit{monopole term} $\varphi^{\left(0\right)}$, which reads\footnote{Of course, $\varphi^{\left(0\right)}$ is dominant as long as the monopole term is nonvanishing.}

\begin{align}
\varphi^{\left(0\right)}\left(\mathbf{r}\right)=\frac{Q}{4\pi r}\left(1-e^{-m_Pr}\right).\label{monopole}
\end{align}

In this equation, $Q$ is the total charge of the distribution:

\begin{align}
Q=\int \rho\left(\mathbf{r}\right)d^3r .\label{total charge}
\end{align}

As we can see, in the monopole approximation, the potential behaves as if all the charge $Q$ were a point charge located at the origin of the coordinate system - see equation (\ref{discrete}) with $n=1$ and $\mathbf{r}_1=\mathbf{0}$.

The next-to-leading order term in (\ref{varphi}) is the so-called \textit{dipole term}:

\begin{align}
\varphi^{\left(1\right)}\left(\mathbf{r}\right)=\frac{1}{4\pi r}\left[\frac{1}{r}-\left(m_P+\frac{1}{r}\right)e^{-m_P r}\right]\mathbf{p}\cdot\hat{\mathbf{r}}.\label{dipole}
\end{align}

Here, mimicking the Maxwellian case, we have defined the \textit{dipole moment} $\mathbf{p}$ as

\begin{align}
\mathbf{p}=\int\rho\left(\mathbf{r}\right)\mathbf{r}\,\,d^3r.
\end{align}

It is interesting to show explicitly the \textit{quadrupole moment} in Podolsky theory as well. It reads

\begin{align}
\varphi^{\left(2\right)}\left(\mathbf{r}\right)=\frac{1}{8\pi r^3}\sum_{i,j=1}^3\hat{r}_iQ_{ij}\hat{r}_j,\label{quadrupole}
\end{align}
where

\begin{align}
Q_{ij}&=\int d^3r'\rho\left(\mathbf{r}'\right)\left[f\left(m_p r\right)r'_i r'_j+g\left(m_p r\right)\left(r'\right)^2\delta_{ij}\right];\\
f\left(x\right)&\equiv 3-e^{-x}\left(x^2+3x+3\right);\\
g\left(x\right)&\equiv e^{-x}\left(x+1\right)-1.
\end{align}

It is worth to note that each and every one of these quantities reduce to their Maxwellian counterpart in the limit $m_P\rightarrow \infty$.


\section{Applications of the multipole expansion}

In this section we apply the techniques developed in the previous sections to some classical problems in Electrostatics and Magnetostatics for the Generalized Theory.


\subsection{The electrostatics of an electric charged disk}

\begin{figure}[h]
\includegraphics[scale=.3]{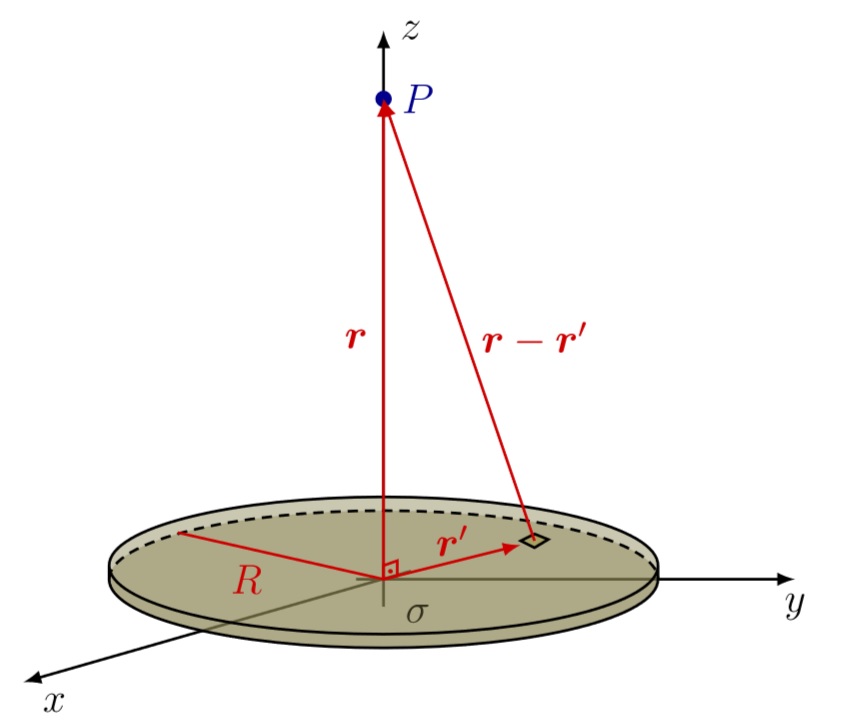}\caption{An electric charged disk \label{fig disk}}
\end{figure}

Let us consider an electric charged disk with negligible thickness and radius $R$. Let us also consider that the surface density of electric charge is constant throughout the disk. Let us call such constant $\sigma$. If we choose the coordinate system properly, we can describe the problem as in the figure (\ref{fig disk}).

We are interested in computing the scalar potential $\varphi$ at the point $P$ whose location is given by the vector $\mathbf{r}$. For simplicity, let us assume the point $P$ lies in the $z$-axis. So, the scalar potential is

\begin{align}
\varphi\left(\mathbf{r}\right)=\frac{1}{4\pi}\int\left(\frac{1-e^{-m_P\left|\mathbf{r}-\mathbf{r}'\right|}}{\left|\mathbf{r}-\mathbf{r}'\right|}\right) \rho\left(\mathbf{r}'\right)d^3r',\label{exact}
\end{align}
where the electric charge density $\rho$ reads\footnote{Here, $\Theta$ is the Heaviside step function defined as $\Theta\left(x\right)=1$ if $x\geq 0$ and $\Theta\left(x\right)=0$ otherwise.}

\begin{align}
\rho\left(\mathbf{r}'\right)= \sigma\Theta\left(R-r'\right)\delta\left(r'_3\right).\label{rho disk}
\end{align}

By truncating the expansion (\ref{varphi}) in its third term, we get the following approximation for the scalar potential:

\begin{align}
\varphi\left(\mathbf{r}\right)\simeq \varphi^{\left(0\right)}\left(\mathbf{r}\right) + \varphi^{\left(1\right)}\left(\mathbf{r}\right)+ \varphi^{\left(2\right)}\left(\mathbf{r}\right).
\end{align}

As it is now usual, $\varphi^{\left(0\right)}$ represents the monopole term, $\varphi^{\left(1\right)}$  the dipole term, while $\varphi^{\left(2\right)}$ is the quadrupole term. Explicit calculations for these terms accordingly to (\ref{monopole}), (\ref{dipole}) and (\ref{quadrupole}) yield:

\begin{align}
\varphi^{\left(0\right)}\left(\mathbf{r}\right) &=\frac{Q}{4\pi r}\left(1-e^{-m_Pr}\right);\\
\varphi^{\left(1\right)}\left(\mathbf{r}\right) &=0;\\
\varphi^{\left(2\right)}\left(\mathbf{r}\right) &= -\frac{Q}{16\pi r}\left[1-e^{-m_Pr}\left(m_Pr +1\right)\right]\left(\frac{R}{r}\right)^2.
\end{align}

Here, following (\ref{total charge}), we have defined the total charge:

\begin{align}
Q=\pi\sigma R^2.
\end{align}

As it happens in Maxwell's case, there is no dipole moment for a uniformly charged disk. Furthermore, the expansion, in this approximation, shows us that the disk can be considered as having all of its electric charge in the origin of the coordinate system, with a small correction due to a quadrupole moment term. The fact that the quadrupole term is small compared to the monopole one is evidenced by the ratio $\varphi^{\left(2\right)}\left(\mathbf{r}\right)/\varphi^{\left(0\right)}\left(\mathbf{r}\right)= \mathcal{O}\left[\left(R/r\right)^2\right]$.

Perhaps one of the most striking features of this example is the fact that it is, indeed, exactly solvable. This allows us to check if the approximation technique developed makes sense. Truly, the exact scalar potential is obtained integrating (\ref{exact}), thanks to the simple form of the electric charge distribution (\ref{rho disk}):

\begin{align}
\varphi\left(\mathbf{r}\right) =&\,\frac{Q}{2\pi R}\left[\left(\frac{r}{R}\right)\sqrt{1+\left(\frac{R}{r}\right)^2}-\left(\frac{r}{R}\right)+\frac{e^{-m_Pr\sqrt{1+\left(\frac{R}{r}\right)^2}}-e^{-m_Pr}}{m_PR}\right].
\end{align}

Now, if we expand this result in powers of $R/r$ and retain only the first terms, we find

\begin{align}
\varphi\left(\mathbf{r}\right) \simeq &\,\frac{Q}{4\pi r}\left(1-e^{-m_Pr}\right)-\frac{Q}{16\pi r}\left[1-e^{-m_Pr}\left(m_Pr+1\right)\right]\left(\frac{R}{r}\right)^2,
\end{align}
which is the same result found in the multipole expansion. Hence, this example corroborates the validity of our multipole expansion for the Podolsky Theory.


\subsection{Magnetostatics of a circular coil with electric current}

In a similar manner as we did for the electrostatic potential, we can write for the vector potential:

\begin{align}
\mathbf{A}\left(\mathbf{r}\right)=\sum_{l=0}^\infty\mathbf{A}^{\left(l\right)}\left(\mathbf{r}\right),\label{A expansion}
\end{align}
where

\begin{align}
\mathbf{A}^{\left(l\right)}\left(\mathbf{r}\right)=\int d^3r'\frac{G_l\left(\mathbf{r},\mathbf{r}'\right)\mathbf{j}\left(\mathbf{r}'\right)}{4\pi r}.
\end{align}

Let us take a look at the first two terms of the multipole expansion for this potential. The first term in (\ref{A expansion}) is the \textit{magnetic monopole}:

\begin{align}
\mathbf{A}^{\left(0\right)}\left(\mathbf{r}\right)&=\left(\frac{1-e^{-m_Pr}}{4\pi r}\right)\int \mathbf{j}\left(\mathbf{r}'\right)d^3r'.
\end{align}

As it happens, $\mathbf{A}^{\left(0\right)}$ can be computed exactly to any (localised) electric current distribution. In order to evaluate it, let us consider the integral of $\nabla\cdot\left[r_i\,\mathbf{j}\left(\mathbf{r}\right)\right]$ over the whole space.\footnote{Here, $r_i$ is the $i$-th component of the vector $\mathbf{r}$.} That volume integral can be converted into a surface integral that vanishes due to the fact that $\mathbf{j}$ is assumed to be localised. From this, it turns out that

\begin{align}
\int j_i\left(\mathbf{r}\right) d^3r &= -\int r_i\nabla\cdot\mathbf{j}\left(\mathbf{r}\right)d^3r.
\end{align}

Now, due to the antisymmetry of the field-strength and the Euler-Lagrange equations (\ref{EL}), we see that the source $J$ satisfies a continuity equation, namely,
$\partial_\mu J^\mu=0$. In the static regime, this continuity equation reads simply

\begin{align}
\nabla\cdot\mathbf{j}\left(\mathbf{r}\right)&=0.
\end{align}

This shows that $\int \mathbf{j}\left(\mathbf{r}\right)d^3r=\mathbf{0}$ and the magnetic monopole term of the multipole expansion vanishes.

Now, let us take a look at the magnetic dipole term:

\begin{align}
\mathbf{A}^{\left(1\right)}\left(\mathbf{r}\right) &=\left[\frac{1-e^{-m_Pr}\left(m_Pr+1\right)}{4\pi}\right]\frac{\mathbf{m}\times\mathbf{r}}{r^3},\label{magnetic dipole}
\end{align}
where $\mathbf{m}$ is the \textit{magnetic dipole moment}

\begin{align}
\mathbf{m} &\equiv\frac{1}{2}\int \mathbf{r}\times\mathbf{j}\left(\mathbf{r}\right) d^3r.
\end{align}

So, in the lowest order of the multipole expansion, the magnetic field has the form

\begin{align}
\mathbf{B}\left(\mathbf{r}\right) \simeq & \,\,\mathbf{B}^{\left(1\right)}\left(\mathbf{r}\right);\\
\mathbf{B}^{\left(1\right)}\left(\mathbf{r}\right) =&\,\frac{1}{4\pi r^3}\left\{\left\{3+e^{-m_Pr}\left[3+3m_P r-\left(m_P r\right)^2\right]\right\}\left(\mathbf{m}\cdot\hat{\mathbf{r}}\right)\hat{\mathbf{r}}\right.\nonumber\\
&\left.-\left\{1+e^{-m_P r}\left[5+5m_P r-\left(m_P r\right)^2\right]\right\}\mathbf{m}\right\},
\end{align}
and we see promptly that the Maxwell's result is obtained in the limit

\begin{align}
\lim_{m_P\rightarrow\infty}\mathbf{B}^{\left(1\right)}\left(\mathbf{r}\right) =&\,\frac{3\left(\mathbf{m}\cdot\hat{\mathbf{r}}\right)\hat{\mathbf{r}}-\mathbf{m}}{4\pi r^3}.
\end{align}

Now, we will consider the example of the multipole expansion approach to a circular coil with electric current.

\begin{figure}[h]
\includegraphics[scale=.3]{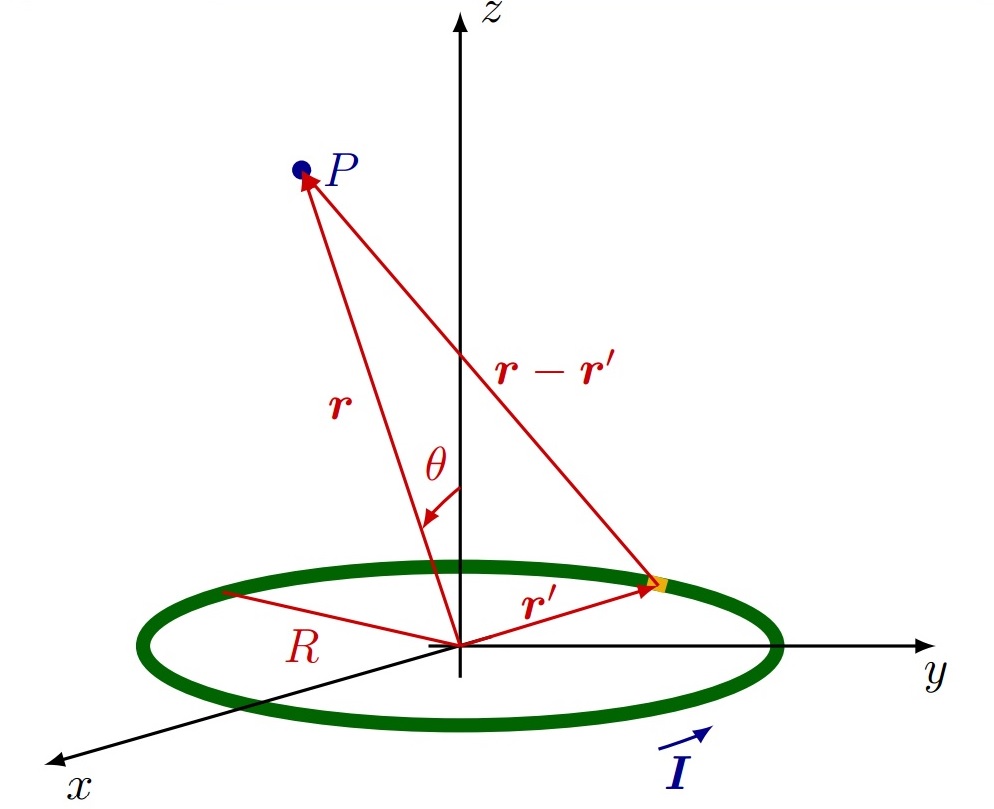}\caption{A circular coil with electric current \label{fig coil}}
\end{figure}

If we consider a circular, conductive, widthless coil with radius $R$ in which it flows an electric current $I$ and we are interested in computing the magnetic field generated by it in a point $P$ not belonging to the coil, we can choose the reference system like in figure (\ref{fig coil}). As indicated in the figure, the electric current $I$ lies in the $x$-$y$ plane and it flows counterclockwise as seen from the positive semi-axis $z$. If the point $P$ is equivalent to the vector $\mathbf{r}$, the vector potential $\mathbf{A}$ at $P$ is

\begin{align}
\mathbf{A}\left(\mathbf{r}\right)=&\, \frac{1}{4\pi}\int G\left(\mathbf{r}-\mathbf{r}'\right)\mathbf{j}\left(\mathbf{r}'\right)d^3r'\label{A in terms of J}
\end{align}
where the current density is written in terms of the spherical coordinates $r$, $\theta$, and $\phi$ as\footnote{As usual, $r\in \left[0,\infty\right)$, $\theta\in \left[0,\pi\right]$, and $\phi\in\left[0,2\pi\right)$.}

\begin{align}
\mathbf{j}\left(\mathbf{r}\right)=&\, \frac{I\delta\left(r-R\right)\delta\left(\theta-\frac{\pi}{2}\right)\hat{\mathbf{\phi}}}{R}.
\end{align}

Here, $I$ is a positive constant. So, the vector potential can be written as

\begin{align}
\mathbf{A}\left(\mathbf{r}\right)=&\,\frac{IR}{\pi h}\left[\left(\frac{2}{k^2}-1\right)\mathbb{K}\left(k\right)-\frac{2}{k^2}\mathbb{E}\left(k\right) +\int_0^{\pi/2}\frac{e^{-m_Ph\mathfrak{h}}}{\mathfrak{h}}\left(1-2\sin^2\gamma\right)d\gamma \right]\hat{\mathbf{\phi}},
\end{align}
where $\mathbb{K}\left(k\right)$ and $\mathbb{E}\left(k\right)$ are the Complete Elliptical Integrals of the First and of the Second Kinds, respectively, and we have defined

\begin{align}
h=&\,\sqrt{r^2+R^2+2rR\sin\theta};\\
k=&\,\frac{\sqrt{4rR\sin\theta}}{h};\\
\mathfrak{h}=&\,\sqrt{1-k^2\sin^2\gamma}.
\end{align}

This is an example of a situation in which we are unable to solve the problem exactly. To circumvent this difficulty, we use our multipole expansion approach. In the lowest order, we can approximate the vector potential $\mathbf{A}$ by its dipole term $\mathbf{A}^{\left(1\right)}$, given by equation (\ref{magnetic dipole}). In the present case, explicit calculations for the magnetic dipole moment yields:

\begin{align}
\mathbf{m}=&\,I\pi R^2\hat{\mathbf{z}}.
\end{align}
Hence

\begin{align}
\mathbf{A}^{\left(1\right)}\left(\mathbf{r}\right)=&\,\frac{IR^2}{4r}\left[1-e^{-m_Pr}\left(m_Pr+1\right)\right]\sin\theta\,\,\hat{\phi}.
\end{align}

The magnetic field, in this approximation, becomes:
\begin{align}
\mathbf{B}\left(\mathbf{r}\right)\simeq &\,\,\mathbf{B}^{\left(1\right)}\left(\mathbf{r}\right);\\
\mathbf{B}^{\left(1\right)}\left(\mathbf{r}\right)=&\,\frac{IR^2}{2r^3}\left\{\left[1-e^{-m_Pr}\left(1+m_Pr\right)\right]\hat{\mathbf{z}} +\left\{3-\left[3+3m_Pr+\left(m_Pr\right)^2\right]e^{-m_Pr}\right\}\frac{\sin\theta\,\,\hat{\theta}}{2}\right\}.
\end{align}
Although this result is rather nontrivial, it satisfies the limit:

\begin{align}
\lim_{m_P\rightarrow\infty}\mathbf{B}^{\left(1\right)}\left(\mathbf{r}\right)=&\,\frac{IR^2}{4r^3}\left(2\hat{\mathbf{z}}+3\sin\theta\,\hat{\theta}\right),
\end{align}
which is exactly the result predicted by Maxwell's theory.

\begin{figure}[h]
\includegraphics[scale=.5]{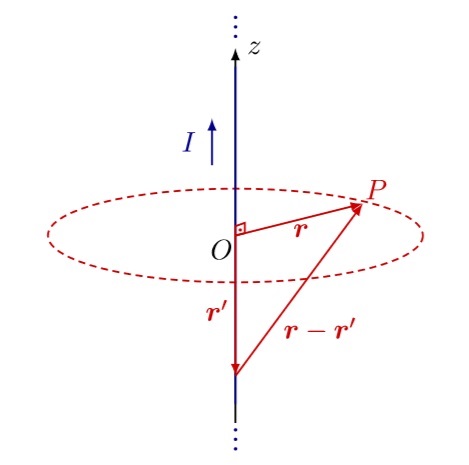}\caption{An electric charged infinite wire \label{fig wire}}
\end{figure}

We would like to close this section with another application in Magnetostatics. One that is exactly solvable. Let us consider an infinite rectilinear wire in which it flows a constant electric current $I$. If the situation is like that of figure (\ref{fig wire}), $\mathbf{j}$ has the form

\begin{align}
\mathbf{j}\left(\mathbf{r}\right)=&\,\frac{I\delta\left(r\right)}{2\pi r}\hat{\mathbf{z}}.
\end{align}

Using this electric current surface density into equation (\ref{A in terms of J}), we can show that the magnetic field takes the form:

\begin{align}
\mathbf{B}\left(\mathbf{r}\right)=&\,\frac{I}{2\pi r}\left[1-2m_PrK_1\left(m_Pr\right)\right]\hat{\phi},\label{B of a wire}
\end{align}
where $K_1$ is the Modified Bessel Function of Second Kind and order 1.

There is a surprising phenomenon encoded in equation (\ref{B of a wire}). In order to reveal it, let us write $\mathbf{B}\left(\mathbf{r}\right)= B\left(\mathbf{r}\right)\hat{\phi}$. As expected, we have $\lim_{m_P\rightarrow\infty}B\left(\mathbf{r}\right)=B_M\left(\mathbf{r}\right)$, where $B_M$ is the Maxwell's result for an infinite wire:

\begin{align}
B_M\left(\mathbf{r}\right)=&\,\frac{I}{2\pi r}.\label{B of Maxwell}
\end{align}

Now, if we compute the limit of small distances in Podolsky's result, we find $\lim_{r\rightarrow 0^+}B\left(\mathbf{r}\right)=-\infty$ (here we are assuming, without loss of generality, $I>0$). However, the correspondent limit of equation (\ref{B of Maxwell}) is $\lim_{r\rightarrow 0^+}B_M\left(\mathbf{r}\right)=+\infty$. This sign difference means that if we are measuring the Podolsky's magnetic field generated by an infinite wire into which a constant electric current flows as a function of the distance, we start with results which are pretty close to those of Maxwell's. Surprisingly, at some point, Podolsky's magnetic field flips sign, and as we are getting closer and closer to the wire, it diverges. It diverges with opposite sign when compared to Maxwell's field, though.

\begin{figure}[h]
\includegraphics[scale=.4]{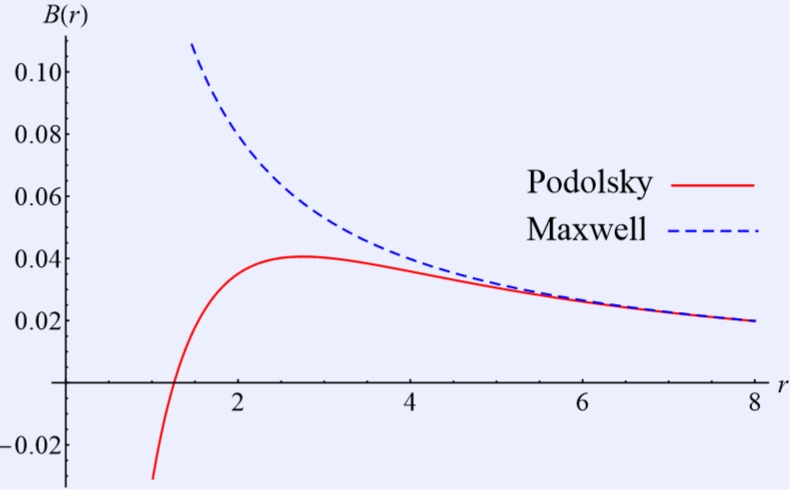}\caption{Podolsky's and Maxwell's magnetic field of an infinite wire \label{fig B}}
\end{figure}

For comparison, Maxwell's and Podolsky's results are plotted in figure (\ref{fig B}).\footnote{In plotting figure (\ref{fig B}), we set $m_P=1$ and $I=1$.} Since Maxwell's field is always positive and for distances large enough Podolsky's field resembles Maxwell's, from the continuity of $B$ as function of $r$, we see that there is a point at which $B$ is maximum and another point at which $B$ vanishes. This latter is the ``flipping point" mentioned earlier. This behavior in the proximity of the wire might be interesting in probing Podolsky theory.

\section{Final remarks}

We have developed the multipole expansion for the Generalized Electrodynamics in the static regime. As we have shown, unlike what happens in Maxwell's theory, in Podolsky's theory the multipole expansion is not straight-forward. Besides the two distance scales already present in Maxwell's expansion, namely, the typical size of the charge (or current) distribution and the distance to the point at which we intend to compute the potential, in Podolsky case appears another physical distance: the Podolsky length $l_P$. With this new length scale, we were able to find a regime of validity for our multipole expansion. As we have shown, this validity consists, basically, in a criterion of truncability for the infinite series. In an analogy to what happens in the usual expansion, we identified the first terms in the expansion as the monopole, the dipole and the quadrupole terms of the potentials and we found a closed form for the $2^l$-pole term of the Green function.

Once we have consistently developed the multipole expansion for the Generalized Theory, we applied our method to some classical problems in electrostatics and magnetostatics. We have studied a charged disk - which is an exactly solvable problem - and we used it as a test to our expansion. As we have shown, all of our results are consistent. So, the charged disk corroborates the validity of our multipole expansion approach. Furthermore, we turned our attention to some applications in magnetostatics. We studied a circular coil - in which we computed explicitly the first terms of the multipole expansion - and an infinite wire, both examples in Podolsky theory. In every single problem studied, we were able to recover Maxwell's results in the proper limit, \textit{i. e.}, when the Podolsky parameter $m_P$ goes to infinity. Surprisingly, the case of an infinite wire, besides being exactly solvable, proved to be very enthralling. As we have seen, the magnetic field generated by an infinite wire (in which it flows an electric current) can be used to test experimentally Podolsky theory. Since we would have to probe the magnetic field very close to the wire in order to see a deviation from Maxwell's prediction, this is a tantalising experimental challenge. For one thing, apparently this cannot be probed with a macroscopic wire. The reason for that is that when gauging the field very close to a macroscopic wire, such a wire cannot be considered as ideal (widthless). So, in the search for deviations of Maxwell's result, a very thin wire is needed.

For future works, we hope to study the radiation theory for Podolsky Electrodynamics and get some insight in the retarded and advanced Green Functions for the theory.

%

CAB thanks FAPESP for supporting the project no. 2011/19306-7, BMP thanks CNPq for partial support, and PHO thanks CAPES for full support.


\end{document}